\documentclass[conference]{IEEEtran}
\IEEEoverridecommandlockouts
\usepackage{cite}
\usepackage{amsmath,amssymb,amsfonts}
\usepackage{multirow}
\usepackage{algorithmic}
\usepackage{graphicx}
\usepackage{textcomp}
\usepackage{xcolor}
\usepackage{flushend}

\def\BibTeX{{\rm B\kern-.05em{\sc i\kern-.025em b}\kern-.08em
    T\kern-.1667em\lower.7ex\hbox{E}\kern-.125emX}}
    
\begin{document}

\title{Clifford Gate Optimisation and T Gate Scheduling: Using Queueing Models for Topological Assemblies}

\author{\IEEEauthorblockN{Alexandru Paler}
\IEEEauthorblockA{\textit{Linz Institute of Technology}\\
\textit{Johannes Kepler University}, Linz, Austria\\
alexandrupaler@gmail.com}
\and
\IEEEauthorblockN{Robert Basmadjian}
\IEEEauthorblockA{\textit{Chair of Sensor Technology}\\
\textit{University of Passau}, Passau, Germany\\
robert.basmadjian@uni-passau.de}
}

\maketitle

\makeatletter
\def\ps@IEEEtitlepagestyle{
  \def\@oddfoot{\mycopyrightnotice}
  \def\@evenfoot{}
}

\def\mycopyrightnotice{
  {\footnotesize
  \begin{minipage}{\textwidth}Copyright \copyright 2019 IEEE.  Personal use of this material is permitted.  Permission from IEEE must be obtained for all other uses, in any current or future media, including reprinting/republishing this material for advertising or promotional purposes, creating new collective works, for resale or redistribution to servers or lists, or reuse of any copyrighted component of this work in other works.\end{minipage}}}

\IEEEpubidadjcol

\begin{abstract}
Clifford gates play a role in the optimisation of Clifford+T circuits. Reducing the count and the depth of Clifford gates, as well as the optimal scheduling of T gates, influence the hardware and the time costs of executing quantum circuits. This work focuses on circuits protected by the surface quantum error-correcting code. The result of compiling a quantum circuit for the surface code is called a topological assembly. We use queuing theory to model a part of the compiled assemblies, evaluate the models, and make the empiric observation that at least for certain Clifford+T circuits (e.g. adders), the assembly's execution time does not increase when the available hardware is restricted. This is an interesting property, because it shows that T gate scheduling and Clifford gate optimisation have the potential to save both hardware and execution time.
\end{abstract}


\section{Introduction}

The efficient design of error-corrected quantum circuits is an open problem, because there is a trade-off between hardware cost and execution time that needs to be accounted for. Very often, for Clifford+T circuits, the goal is to optimise T-count and/or T-depth. However, compiling and optimising error-corrected quantum circuits focuses on more than just the T gates. This work presents preliminary evidence supporting the goal of Clifford gate optimisation. We show that such optimisation methods can have a significant influence on the overheads of error-corrected quantum circuit execution.

\subsection{Surface Code Assembly Volume}

\begin{figure}[t!]
    \centering
    \includegraphics[width=0.6\columnwidth]{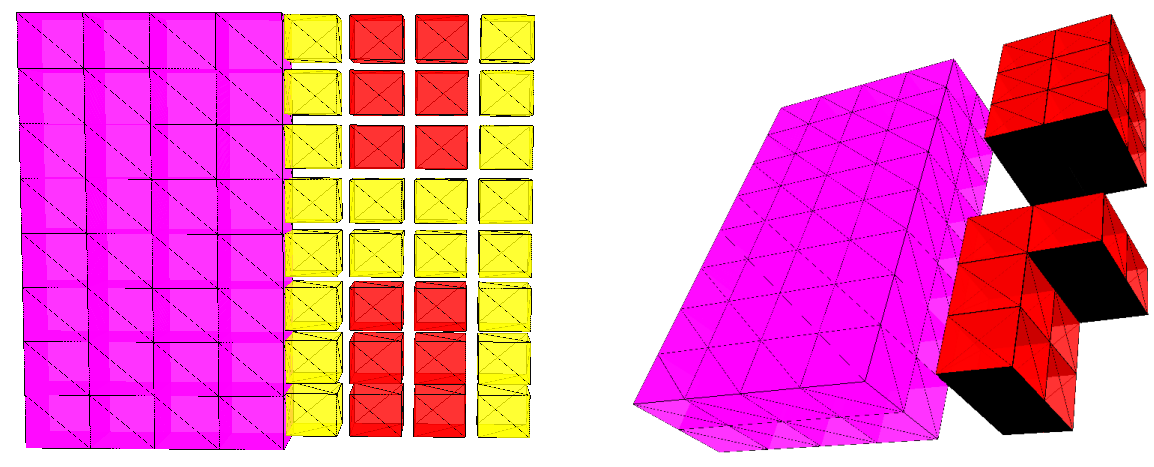}
    \caption{3D view of a topological assembly with two layers. (a) A region is for distillations (magenta), data patches are in columns (red), ancilla patches (yellow) are between the data patches. (b) Patches were placed in two layers. The ancilla are not used, and the distillation procedure is being executed. Execution time flows from image background to foreground.}
    \label{fig:assembly}
\end{figure}

The surface quantum error-correcting code (QECC) is at the foundation of the most promising quantum computer architectures. The code tolerates high hardware error rates, and has straightforward architectural requirements \cite{fowler2012surface}. A \emph{topological assembly} (e.g. Fig.~\ref{fig:assembly}) is the result of compiling a quantum circuit to surface QECC elements. Compilation requires the circuit being transformed into Clifford+T gates. The surface code operates at a so-called logical layer, and the physical layer can be for example a topological cluster state (3D variant of the code), or a lattice of physical qubits arranged according to well defined 2D nearest neighbour interactions (planar surface code).

\begin{figure}[t!]
    \centering
    \includegraphics[width=0.6\columnwidth]{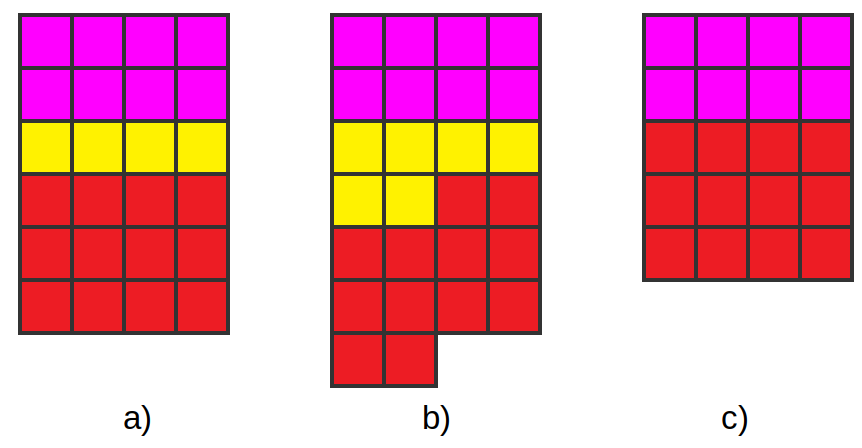}
    \caption{Assembly layer. Distillery region is magenta. The yellow region stores distilled states (e.g. capacity 4). Clifford+T computation is executed using buffered T states in the red region (e.g. 12 logical qubits). Considering magenta and red regions of fixed sizes, the number of physical qubits is reduced by decreasing the size of the yellow region.}
    \label{fig:2}
\end{figure}

There are different surface code techniques (e.g. braiding, lattice surgery \cite{horsman2012surface}). The discussion herein, and the accompanying figures, are based on the assumption that the code is implemented through lattice surgery. Nevertheless, our work is valid for the braided version of the code, too. In lattice surgery, logical qubits are encoded into code patches (e.g. squares in Fig.~\ref{fig:2}), which have two boundary types (opposite boundaries are of the same type). A quantum computation is executed by choosing on which boundary to interact patches (merged and split). Patch dimensions are dictated by the distance of the surface code.

The overheads (costs) of applying the surface code (in fact any QECC) is quantified by the utilised number of physical qubits, and the generated time overhead. The goal of fault-tolerant quantum circuit optimisation is to reduce the overheads (hardware and time) without sacrificing the strength of the error-correction. The overhead of a circuit's assembly can be abstracted through a 3D spacetime volume (qubits $\times$ time). In 3D, one of the dimensions is abstracted time. For the lattice surgery planar code, space (hardware, cf. Fig.~\ref{fig:2}) is the two-dimensional area necessary for storing all the logical qubit patches.

\subsection{Distillation}

The Clifford gates can be implemented fault-tolerantly in the surface code, but the T gates require the support of high fidelity T states. Such states need to be distilled first. A distillation procedure is a sub-circuit which takes multiple low-fidelity instances of the same state and outputs probabilistically a higher-fidelity state -- a distilled state. The complexity of implementing surface QECCs originates from the necessary \emph{distillation} procedures\cite{fowler2012surface}. Executing a T gate comes at the additional cost of executing a distillation, which in turn is a computation with relatively high assembly volume. It is an open question how to design low volume distillations \cite{ding2018magic}.

Each distillation has a known duration in time, which is assumed to be equal to a multiple of the code distance. The manipulation of patches in time is represented by a cube, and a distillation is a large cuboid whose volume equals multiple cubic volume units (e.g. in Fig.~\ref{fig:assembly} the magenta cubes are volume units of a \emph{single} distillation that was executed for two time steps, and will finish after additional steps, which are not illustrated).

\begin{figure}[h!]
    \centering
    \includegraphics[width=0.9\columnwidth]{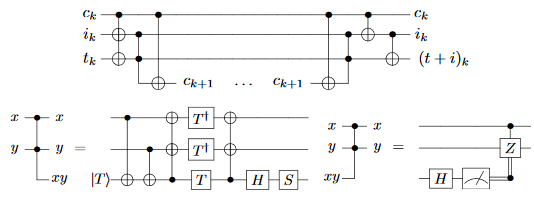}
    \caption{The considered circuits are $n$-qubit carry ripple adders. Due to how gates are decomposed into Clifford+T, the resulting circuits include $4n$ T gates. All the T gates appear in the first half of the circuit. However, due to the carry ripple structure, some T gates can be delayed without delaying the overall computation.}
    \label{fig:adder}
\end{figure}

\subsection{Worst-Case: T-count equals T-depth}

The current generation of quantum machines (almost certain the next generation, too) are resource restricted, and will support only sequential distillations (parallel would require too many physical qubits). This effectively calls for the adaptation of the Clifford+T circuits: the initially parallel $T$ gates (e.g. Fig.~\ref{fig:adder}) have to be \emph{scheduled such that their sequential execution introduces a minimal time overhead into the assemblies}.

In hardware restricted scenarios it is assumed that the worst case execution time of a circuit is bounded by the T-count (when all T gates are executed sequentially T-depth equals T-count). In general, an assembly's depth (time, cf. \emph{circuit depth}) is a function of the number of Clifford+T gates in the circuit. Worst-case estimations of assembly volumes use, as an indication for depth, the circuits' T-count (the number of T gates in the circuit) \cite{gidney2018efficient}.

\subsection{Problem Statement}

The following definitions are used to state the problem. Considering Fig.~\ref{fig:assembly}, a \emph{layer} (e.g. Fig. \ref{fig:2}) is a slice of the assembly made at a given time coordinate. A \emph{distillery} is the assembly region (cf. magenta in Fig. \ref{fig:2}) where distillations are executed. T gates are executed when a distillation succeeded, and a distilled T state was obtained. It is reasonable to store distilled states into a \emph{buffer}. Buffering distilled T states has the potential of shortening the depth of the topological assembly, because T gates would be parallelised. However, this comes at the expense of hardware overheads (buffer sizes).

Patches (red or yellow) are interacted by manipulating their boundaries (merges and splits). In Fig.~\ref{fig:assembly}, red patches represent logical qubits, and yellow patches are ancillae (used to intermediate between red patches). This work focuses on distilled states, and we consider that the yellow patches represent distilled T states, and that computational ancillae are red. This graphical simplification (cf. Figs.~\ref{fig:assembly} and \ref{fig:2}) does not affect the generality and conclusion of the analysis. Thus, the yellow region is a visualisation of the buffer. The number of yellow patches in the region indicates the buffer size.

How can the trade-off between hardware and time be explained through the layout of the topological assembly? Does the buffer size influence the execution time? The same question can be reformulated as: Is T-depth always the worst case? This work will provide evidence that for some circuits the assembly depth is larger than T-depth.

\section{Methods}

We present a methodology based on a straightforward queuing model of the distilled states buffer. The methodology is general and applicable to any Clifford+T quantum circuit. We use quantum adders to collect empirical evidence that scheduling T gates, as well as smart distillery control mechanisms lead to computational resource savings. The analysis will show that buffers are not necessary as long as \emph{T gates are perfectly scheduled (on average, one T gate per distillation execution)}. If perfect scheduling is not possible, then the proposed methodology is applicable to:
\begin{itemize}
    \item determine an optimal buffer size;
    \item compute the optimal point in time to turn off a distillery, and use the extra available space to perform the Clifford gate optimisation.
\end{itemize}

The later application implies that the distillery can be stopped before the buffer reaches full capacity, and the available space (yellow patches) is used to speed-up the Clifford part, such that the next T gate will be executed sooner. Thus, the queue based methodology offers quantitative insights into the necessity of future work on Clifford gate optimisation and T gate scheduling.

\subsection{Related work}

Although not explicitly stated, the need for Clifford gate optimisation was observed in \cite{amy2016estimating}. The therein analysed circuits were Clifford dominated, and the T-count was not the upper bound of the worst case resource estimation. Clifford dominated circuits appeared in \cite{babbush2018encoding}, too. Because of their structure, efficient scheduling of T gates became a necessity: distribute (commute where possible) T gates, such that there is no need for buffering distilled T states -- the buffers would have occupied hardware resources. T gate scheduling implied packing Clifford gates between the sequentially executed T gates (see worst case above).

The need for T gate scheduling resulted in the need of controlling distilleries, and the first investigations have been presented in \cite{paler2018controlling}. Therein, the buffer had a fixed size, and could store seven distilled states. Once the buffer was full, the distillery was stopped (no distillations were started until the first buffered state was not consumed by a T gate). Otherwise, the buffer would have overflown. In that work a simplistic look-ahead strategy for calculating the number of next T gates, and necessary distilled T states, was presented. That strategy was used to predict when the buffer would be full.

Nevertheless, previous work has not answered the following questions: 1) are T state buffers always needed? In case they are, what will be the optimal buffer size? How can the buffer size be computed? Can the need for buffering be circumvented? In \cite{paler2018controlling}, once the buffer capacity was capped, the distillery control did not generate additional time overheads in the resulting assemblies.

\subsection{Distillery and Buffer}

\label{sec:model}
\begin{figure}
    \centering
    \includegraphics[width=\columnwidth]{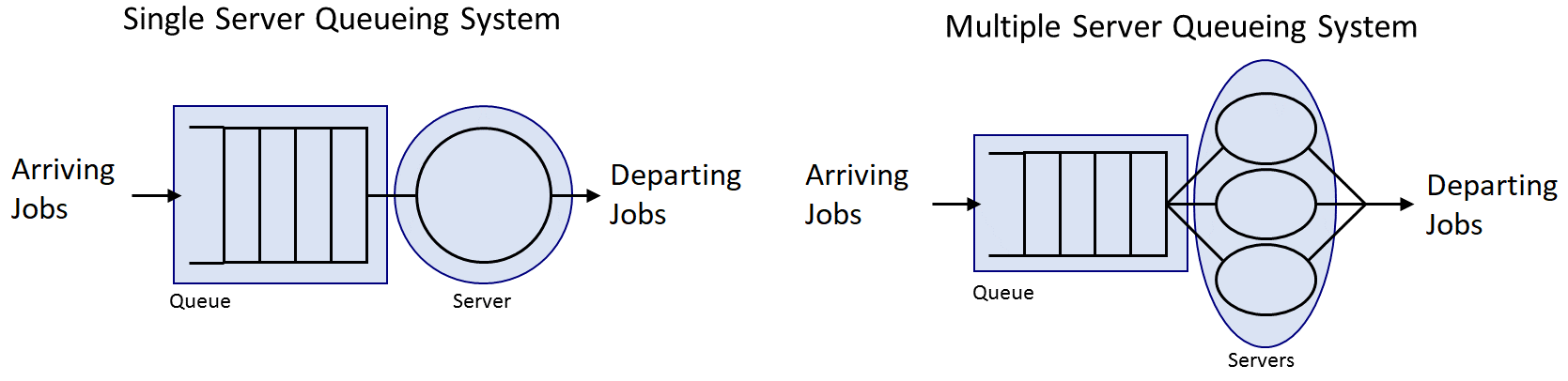}
    \caption{An example of a single and multiple servers queuing systems.}
    \label{fig:queuingsystem}
\end{figure}

We introduce an analogy, based on queuing theory, for capturing how the management of distillation procedures influences the arrangement of patches in the assembly. The analogy will help answer the trade-off question by looking at buffer occupancy. 

In the case of queuing systems, \emph{jobs} (also known as requests, customers, or calls) are stored temporarily in a \emph{buffer} (also known as queue) until the server finishes executing the current job. Thus, a server can execute one job at a time. Buffers can be classified based on their capacity: \emph{finite} or \emph{infinite}. Unlike infinite buffer capacity, the finite one has a limited buffer size.   
 
A \emph{server} is a job processing unit. Upon finishing the execution of a job, a server retrieves the next job from the buffer based on a given \emph{strategy} (e.g. FCFS -- first come, first serve), and executes the job. Parallel job execution is achieved through multiple servers queuing system. Thus, a queuing system consists of one buffer and at least a single server. Figure \ref{fig:queuingsystem} illustrates a graphical presentation of queuing systems consisting of a single and multiple servers. This work considers simple systems with one buffer and one server (e.g. single server queuing system).

In the case of topological assemblies, the job analogue is the \emph{distilled T state}. The buffer's analogue is the \emph{buffer region}. The capacity is the number of distilled states that fit into the buffer. Thus, infinite capacities are not realistic. The equivalent of the server is the region that consumes distilled T states.

Whenever the buffer reaches its full capacity (e.g. overflow), then either the new arriving job is lost or the job  producer (e.g. distillery) is notified about not to send new jobs until there is enough available capacity. In this paper, if such a buffer overflow  happens, we consider the second option of stopping the production of new jobs by the distillery.

\begin{figure}[t!]
    \centering
    \includegraphics[width=0.7\columnwidth]{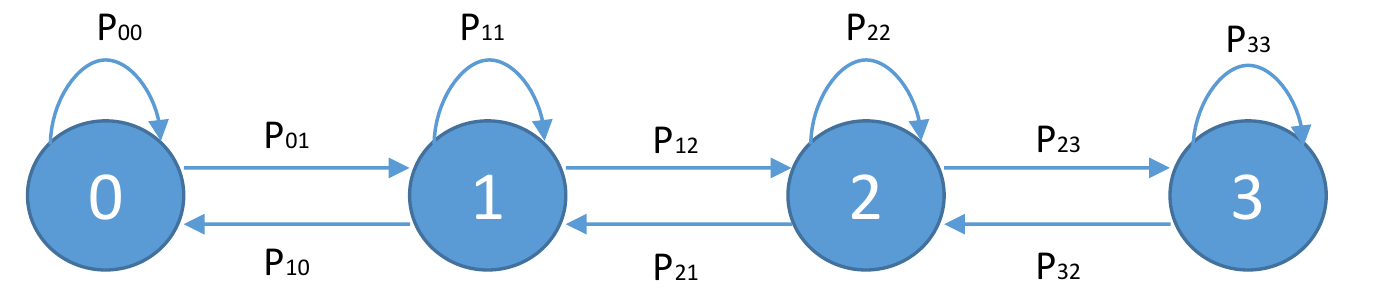}
    \caption{DTMC for distillery buffer capacity.}
    \label{fig:my_label}
\end{figure}

We propose the following methodology to identify the optimal buffer size for a given circuit: 1) the yellow region is modelled as an infinite buffer; 2) the behaviour of the distillery is emulated with maximum one distillation at a time; 2a) buffer occupancy increases by maximum one at a time, because of the sequential distillation process; 2b) buffer occupancy decreases by maximum one at a time, because circuit gates are not parallel (in this work).

The methodological steps are based on Markov chains, where each state is \emph{the number of jobs in the system} (the ones kept in the buffer, and the one used for T gate). This adoption is valid since our problem satisfies the memory-less property: \emph{future depends only on the current state and not: (1) on the history of states, and (2) how much time is spent in the current state}. The topological assembly has a discrete structure along the time axis, and this leads to the formulation of Discrete Time Markov Chains (DTMC, cf. Fig.~\ref{fig:my_label}): the state-space (e.g. $0, 1, 2, \ldots, n$ jobs) and the time-space (e.g. $1, 2, \ldots, n$ time instance) are discrete.

The resulting DTMCs are \emph{ergodic} because (1) these have a finite number of states $N$, (2) their states are pairwise reachable, (3) the chain is aperiodic (e.g. no periodicity of a certain state exist). The ergodicity of a given DTMC ensures the fact that there exists a unique steady-state probability. 

\subsection{Buffer Size, Number of Transitions}

\begin{figure}[t!]
    \centering
    \includegraphics[width=0.9\columnwidth]{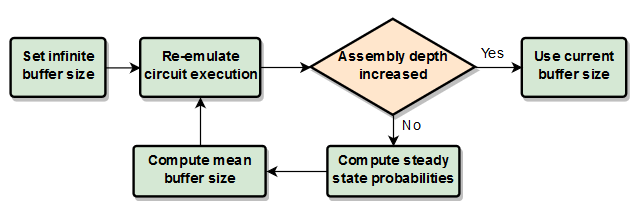}
    \caption{The process of determining optimal buffer size.}
    \label{fig:method}
\end{figure}

Quantum hardware is not an abundant resource, and we analyse the trade-off between optimal buffer capacity, denoted by $B$, and the time overhead. This is an iterative process (cf. Fig.~\ref{fig:method}) in which circuit execution is re-emulated for values in the range $b \in [0, B)$, which will lead to \emph{different number of transitions and different transition probabilities} (incorporating the resulting time overheads, e.g. increased assembly depth). \emph{The worst-case execution time is obtained for the most hardware efficient setting when $b=0$ (e.g there is no buffer).}

The values of transition probabilities $P_{i,j}$ (in matrix form) are obtained through a process which we call \emph{(re)-emulation} (straightforward, linear procedure in which the usage of the buffer is simulated, as distilled states are produced, stored and used). Emulation is not quantum circuit execution or simulation. The methodology iterates the emulation of the corresponding circuit for different number of qubits for infinite and fixed (e.g. 7, see next Section) buffer capacities.

The $P_{i,j}$ transition probabilities are used to compute the \emph{unique steady-state probabilities}, based on the classic system of equations  $\vec{\nu}=\vec{\nu} P$, such that $\sum_{\forall i}{\nu_i} = 1 $ (\cite[p. 41]{bolch}). The transition probability $P_{i,j}$ denotes the probability of reaching state $j$ from state $i$.

Once those steady-state probabilities are calculated \cite{Basmadjian:ondemand,Basmadjian_DVFS_conservative}, it is possible to compute among other the following metrics, which are at the foundation of evaluating the performance of a queuing system: 
\begin{enumerate}
    \item the probability of the system being idle $v_0$ or full $v_n $,
    \item mean number of jobs in the system $\bar{K} = \sum_{i=1}^{\infty}k\cdot v_i$, where $k$ denotes the current state,
    \item the average utilisation of the whole system $\bar{U}=1-v_0$, where $v_0$ is the steady-state probability of an idle system.
\end{enumerate}

\begin{table}[t!]
 \centering
 \caption{Results for infinite and size-7 buffers.}
  \label{tbl:results}
  \footnotesize
  \begin{tabular}{|c|c|c|c|c|c|}
    \hline
    \multirow{2}{*}{Qubit} &
      \multicolumn{2}{c|}{Mean \# of Jobs} &
      \multicolumn{1}{c|}{Total \# of } &
      \multicolumn{1}{c|}{Utilisation } &
      \multicolumn{1}{c|}{Number of } \\
   & Size-7 & Infinite &States (Infinite)& &  Transitions  \\
    \hline
    16 & 2.80 & 2.96 &9& 69\% &  270  \\
    \hline
    32 & 3.85 & 6.51 &19& 73\%  &  558 \\
    \hline
    64 & 4.35 & 13.61 &37& 76\%  &  1134 \\
    \hline
    128 & 4.59 & 27.83 &73& 77\%  & 2286  \\
    \hline
    256 & 4.71 & 56.28 &147& 77\%  & 4590  \\
    \hline
    512 & 4.77 &113.17  &293 & 78\% & 9198  \\
    \hline
    1024 & 4.80 & 226.94 &585 & 78\% &  18414 \\
    \hline
    1536 & 4.80 & 340.72 &878 & 78\% & 27630 \\
    \hline
    2048 & 4.82& 454.5 &1171 & 78\%  & 36846 \\
    \hline
  \end{tabular}
\end{table}

\section{Result: Buffering does not shorten depth}

The influence of the buffer size on practical and commonly used quantum addition circuits \cite{gidney2018halving} is presented in Table~\ref{tbl:results}. It should be noted that, as mentioned in the caption of Fig.~\ref{fig:adder}, the distribution of T gates in the original adder is not perfect. \emph{The circuits we analysed in this work have been previously partially optimised by hand with respect to their T gate scheduling. Nevertheless, it was not obvious from their structure what the optimal buffer capacity will be, or what their maximum assembly depth is}.

To evaluate and analyse the impact of buffer sizes, we setup two configurations for the queuing system: finite buffer size of 7, and infinite capacity. The raw information about distillation behaviour and buffer usage were calculated using SurfBraid\footnote{https://alexandrupaler.github.io/quantjs/}, and a customised Python tool\footnote{https://github.com/alexandrupaler/distilleryqueue} to compute the steady states and the average buffer sizes.

Both cases (finite and infinite) have the same total number of transitions as well as utilisation rate. Regarding the number of transitions: it almost doubles with each doubling of qubit numbers. As circuits get larger, these include more gates. The surprising observation is that irrespective of the buffer size, the execution time of the adders stays the same for the same number of qubits -- no delays by executing sequential T gates. This same doubling behaviour is also evident for the total number of states in the case of infinite buffer capacity. Note that for the case of finite buffer size, this is not happening because the maximum number of states equals buffer capacity plus one (from 0 to 8).

Regarding utilisation, it is almost constant and ranging between 70\% and 78\% both for finite and infinite buffer size independent of the qubits. This indicates that the buffer is on average  empty for around 20\% of the time. Furthermore, the probability that the buffer is full (e.g. $v_n$) never passed 0.05 (e.g. for 16 qubit adder). This denotes that the system never reached to a buffer overflow situation.

By further investigating the mean number of jobs, which is calculated based on steady-state probabilities (omitted from Table~\ref{tbl:results}), the obtained results show that for a finite buffer size of 7, the average was almost constant for all circuit sizes (e.g. ranging from 2.8 till 4.8) for any number of considered qubit. This further motivates the fact that for the considered circuit there is no need to allocate large buffer sizes.

Using the methodology from Fig.~\ref{fig:method}, we reached the conclusion that, for the investigated adders, no buffer is in fact necessary because: a) no buffer overflows were recorded while re-emulation the circuit, b) the assembly depth did not increase even for capacity zero buffers. Consequently, the hardware requirements from buffer perspective can be reduced and optimised adequately.


\section{Conclusion}

The analysed adder has the property that the sequential execution of T gates does not influence the depth of the resulting assembly (execution time). We conjecture that this property is because of the advantageous scheduling of the T gates inside the adders: \emph{the distillery produces states faster than the main computation is consuming them}.

Another interesting property of the adder is that its \emph{depth is not dominated by the T-count}. The assembly depth is larger than the one computed by looking at the T-count. Thus, \emph{Clifford gate optimisation} is a realistic option for circuit optimisation. Future work directions are:
\begin{enumerate}
    \item Optimal scheduling methods for T gates;
    \item Optimisation of the Clifford part (reduce average number of Clifford gates between two subsequent T gates $\equiv$ adapt to distillery speed);
    \item Queuing-based models to assist in the automatic design and analysis of the resulting assemblies.
\end{enumerate}

\section*{Acknowledgements}
AP acknowledges support from the Linz Institute of Technology project CHARON, and a Google Faculty Research Award.

%
\tiny
\bibliographystyle{unsrt}
\bibliography{paper}

\end{document}